\documentstyle[amssymb,twocolumn,prl,aps,epsfig]{revtex}

\twocolumn

\begin{document}
\draft

\wideabs{

\title{Comment on ''$"$Forbidden$"$ transitions between quantum Hall and insulating phases in p-SiGe heterostructures''}

\author{S. S. Murzin}

\address{Institute of Solid State Physics RAS, 142432, Chernogolovka, Moscow District, Russia}
\maketitle

\begin{abstract}

~~~It is shown that recent\cite{Krav} and earlier \cite{Pud1,Pud6,Fang}
experiments, which claimed to observe a disagreement with the global phase diagram
(GPD)\cite{KLZ} of the quantum Hall effect, do not, in fact, contradict the
GPD. Two aspects should be taken into account: (i) insulating phases between quantum 
Hall phases are possible owing to the Shubnikov-de Haas oscillations of the ''bare'' 
diagonal resistivity $\rho _{xx}^{0}$; (ii) according to the two-parameter scaling 
theory\cite{Khm1,Pr}, the filling factor $\nu$ does not determine directly the 
positions of the quantum Hall phases on the magnetic field axis at 
$\omega _{c}\tau \lesssim 1$.

\end{abstract}

}

In a recent paper\cite{Krav} Sakr {\em et al.} reported that they observed
multiple quantum Hall-insulator-quantum Hall transitions (MQH-I-QHT) in
p-SiGe heterostructures, where insulating phases occurred between quantum
Hall (QH) phases at filling factors $\nu =1$ and $2$, $2$ and $3$, and $4$
and $6$. Previously it was reported that MQH-I-QHT had been observed in
silicon MOSFETs\cite{Pud1,Pud6}, and the insulating phase had been detected
between QH phases at $\nu \approx 1.5$ \ in p-SiGe heterostructures\cite
{Fang}. The authors of the papers \cite{Krav,Pud1,Pud6,Fang} claimed that
their results are in a glaring contradiction with the global phase diagram
(GPD) for the QH effect \cite{KLZ}, which follows from the scaling theory 
\cite{Khm1,Pr}, and a new type of the phase diagram was suggested\cite
{Krav,Pud6}.

In this comment I argue that, in fact, the experimental results\cite
{Krav,Pud1,Pud6,Fang} do not contradict the GPD if two circumstances are
taken into account:

(i) The insulating phase between the QH phases is possible owing to the
Shubnikov-de Haas oscillations of the ''bare'' diagonal resistivity $\rho
_{xx}^{0}$, which corresponds to diffusive motion of electrons without
interference effects over a distance larger than diffusion length.

(ii) According to the two-parameter scaling theory\cite{Khm1,Pr}, the
filling factor $\nu =nh/eB$ does not determine directly positions of the QH
phases on the magnetic field axis at $\omega _{c}\tau \lesssim 1$ ($\omega
_{c}=eB/m$ is the cyclotron frequency, $\tau $ is the transport relaxation
time).

The scaling theory presented graphically by the flow diagram\cite{Khm1,Pr}
deals with the Hall $\sigma _{xy}$ and the diagonal $\sigma _{xx}$
conductivity components, and it does not require the Landau quantization of
the electron spectrum. The Landau quantization is incorporated into the
theory through the starting values for the renormalization (i.e. the change
in the conductivity due to diffusive interference effects) which are the
''bare'' conductivities $\sigma _{xy}^{0}$ and $\sigma _{xx}^{0}$ \cite{Pr}.

For a totally spin polarized electron system, maxima of the Shubnikov-de
Haas oscillations of the ''bare'' resistivity $\rho _{xx}^{0}$ occur when
the centers of \ the Landau levels 
\begin{equation}
E_{i}^{L}=(i+1/2)\hbar \omega _{c}\equiv (i+1/2)\frac{E_{F}}{\nu }
\label{El}
\end{equation}
cross the Fermi level $E_{i}^{L}=E_{F}$ (as is shown in Fig.1 for the case $%
E_{F}=const$) at half-integer filling factors $\nu =i+1/2$. Here $i$ is an
integer. The 

\bigskip \epsfig{figure=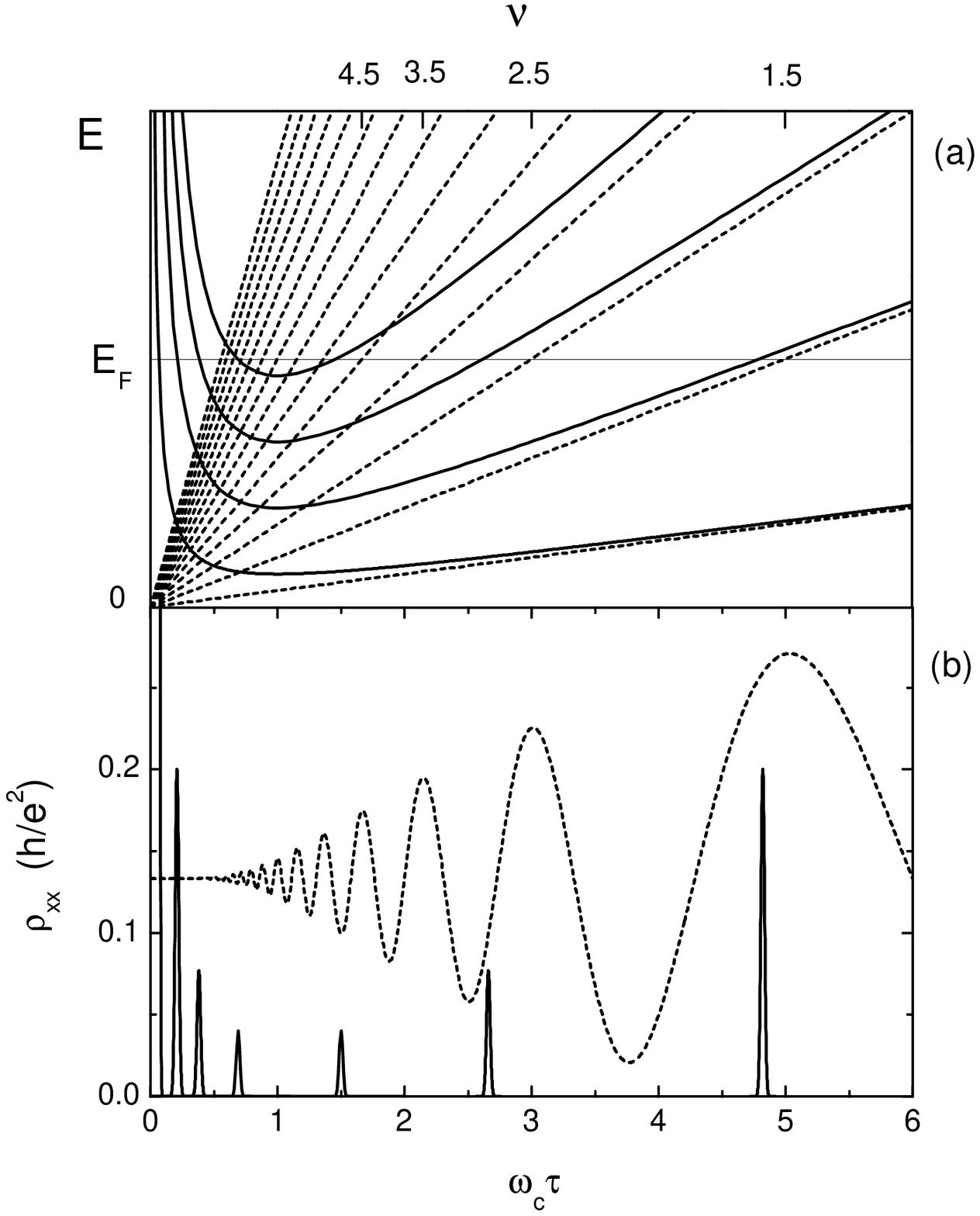,width=8cm,clip=} 
\begin{figure}[tbp]
\caption{Sketch of the magnetic-field dependence (a) of energies of the
Landau levels $E_{i}^{L}$ (dashed line) and of critical states $E_{i}^{c}$
(solid lines), and (b) of the ''bare'' diagonal resistivity $\protect\rho %
_{xx}^{0}$ (dashed lines) and low temperature resistivity of $\protect\rho %
_{xx}$ corresponding to the quantum-Hall-effect regime (solid lines) for the
case of the totally spin-polarized 2D electron system. The horizontal solid
line in the top figure plots the Fermi level $E_{F}$.}
\label{lev}
\end{figure}

~~\epsfig{figure=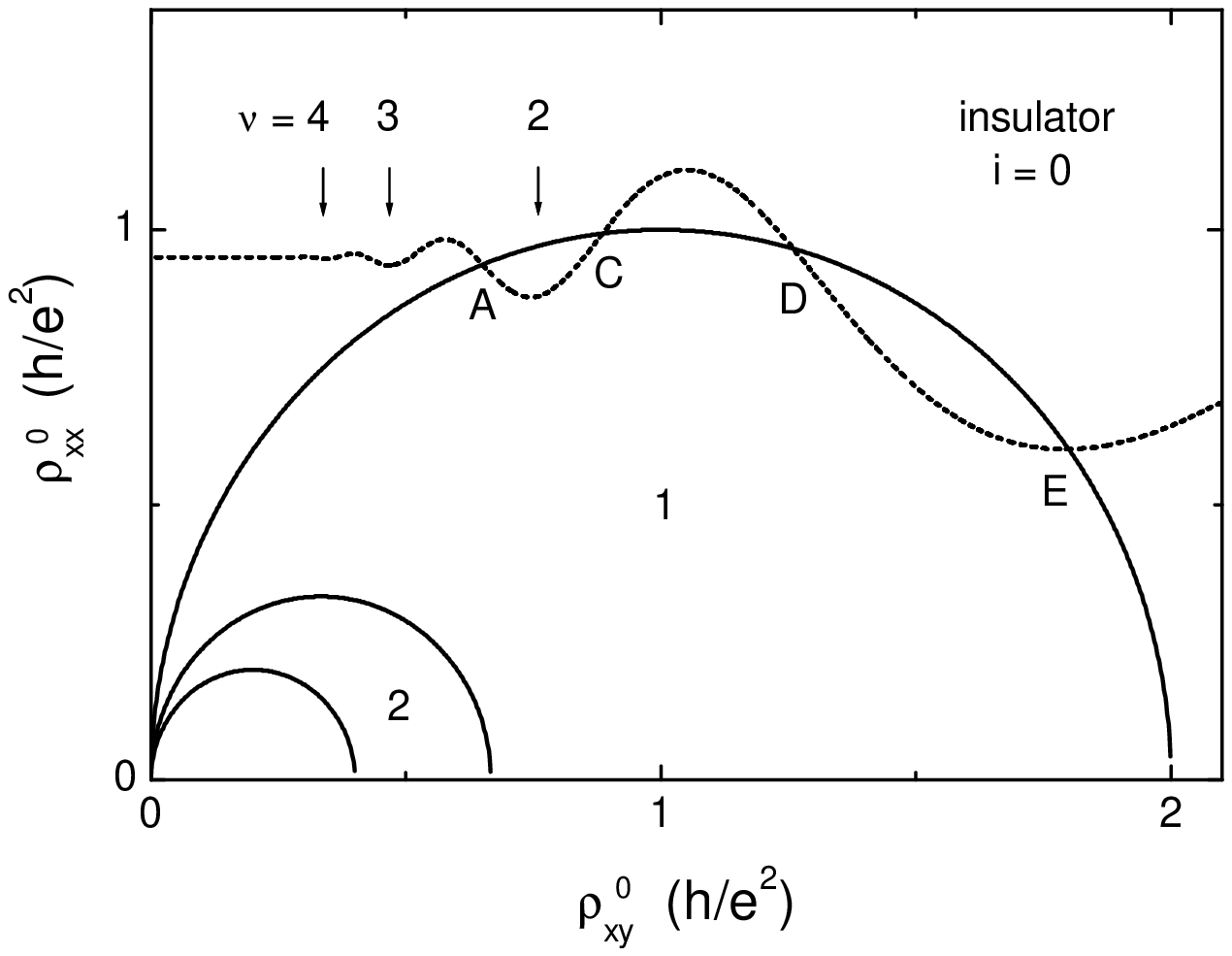,width=7cm,clip=} 
\begin{figure}[h]
\caption{Sketch of the phase diagram of the quantum Hall effect (solid
lines) and the curve of $\protect\rho _{xx}^{0}(\protect\rho _{xy}^{0})$
(dashed line) for the case of the totally spin-polarized 2D electron system.}
\label{GPD}
\end{figure}
renormalization leads to a transformation of the magnetic-field dependence
of $\rho _{xx}^{0}$ with the Shubnikov-de Haas oscillations to the QH effect
picture with peaks in the resistivity $\rho _{xx}$ (and conductivity $\sigma
_{xx}$) separating different QH phases at magnetic fields where

\begin{equation}
\sigma _{xy}^{0}=(i+1/2)e^{2}/h.  \label{Gxy}
\end{equation}
The Hall conductivity $\sigma _{xy}=\sigma _{xy}^{0}$ is not renormalized at
these fields. The nonzero $\sigma _{xx}$ implies that the extended states
are situated at the Fermi level.

The energies of the extended states, $E_{i}^{c}$, as functions of magnetic
field can be calculated by Eq.(\ref{Gxy}) with the classical expression for $%
\sigma _{xy}^{0}$%
\begin{equation}
\sigma _{xy}^{0}(E)=\frac{n(E)e^{2}\tau }{m}\frac{\omega _{c}\tau }{1+\left(
\omega _{c}\tau \right) ^{2}}=\frac{e^{2}}{h}\frac{E\tau }{\hbar }\frac{%
\omega _{c}\tau }{1+\left( \omega _{c}\tau \right) ^{2}}.  \label{cl}
\end{equation}
Here the conduction band is occupied up to energy $E$, $n(E)=Em/h\hbar $ is
the electron density, $\hbar =h/2\pi $. As a result\cite{Khm,Laugh}

\begin{equation}
E_{i}^{c}=(i+1/2)\hbar \omega _{c}\left[ 1+\frac{1}{\left( \omega _{c}\tau
\right) ^{2}}\right]  \label{Ec}
\end{equation}
(see Fig.1). The positions of the $\rho _{xx}$ peaks are determined by
equation $E_{i}^{c}=E_{F}$. It is assumed here that $\tau $ is independent
on the energy $E$. This assumption, however, is not important for finding
the peaks positions since only the value of $\tau $ at the Fermi level is
essential for this calculation.

For a pure 2D system at high magnetic field, when $\omega _{c}\tau \gg 1$,
the filling factor $\nu =nh/eB$ is equal to $\sigma _{xy}^{0}h/e^{2}$.
Therefore, peaks of $\rho _{xx}$ should occur at the same fields where the
Shubnikov-de Haas oscillations have maxima. For $\omega _{c}\tau \lesssim 1$%
the value $\sigma _{xy}^{0}h/e^{2}$\ is different from $\nu $ and the
positions of the $\rho _{xx}$ peaks are very different from the positions of
the maxima of the Shubnikov-de Haas oscillations.

In the plane $\rho _{xy}^{0}-\rho _{xx}^{0}$ the phase boundaries are
semicircles (see Fig.2) described by the equation\cite{KLZ} 
\begin{equation}
\sigma _{xy}^{0}=\frac{\rho _{xy}^{0}}{\left( \rho _{xx}^{0}\right)
^{2}+\left( \rho _{xy}^{0}\right) ^{2}}=(i+1/2)e^{2}/h,  \label{rr}
\end{equation}
following from Eq.(\ref{Gxy}). At a given magnetic field, the quantized
value of the Hall resistance $\rho _{xy}=h/ie^{2}$ is determined by the
position of point $(\rho _{xx}^{0},\rho _{xy}^{0})$ on the phase diagram.
For example, for $(\rho _{xx}^{0},\rho _{xy}^{0})$ located between the two
upper semicircles, $\rho _{xy}=1$. As the magnetic field increases, $\rho
_{xx}^{0}(\rho _{xy}^{0})$ follows a straight horizontal line which crosses
each phase boundary two times because $\rho _{xx}^{0}=m/ne^{2}\tau $ is
independent of the magnetic field. The QH phases are situated between low-
and high-field insulators in this model.

The classical expression (\ref{cl}) for $\sigma _{xy}^{0}$ is quite adequate
at low magnetic fields, $\omega _{c}\tau \ll 1$, and at high magnetic field, 
$\omega _{c}\tau \gg 1$. At $\omega _{c}\tau \sim 1$\ the Shubnikov-de Haas
oscillations of $\sigma _{xy}^{0}$ should be taken into account in accurate
calculations of $E_{i}^{c}$. This could give rise to additional crossing
points between lines $E_{i}^{c}(\omega _{c}\tau )$ and the Fermi level,
hence we have more crossing points between the curve of $\rho _{xx}^{0}(\rho
_{xy}^{0})$ and the phase boundary (see Fig.2), as compared with the case of
the classical expression for $\sigma _{xy}^{0}$. The insulating phase occurs
between the QH phases at $\rho _{xy}^{0}(C)<\rho _{xy}^{0}<\rho _{xy}^{0}(D)$%
, in addition to low- $\rho _{xy}^{0}<\rho _{xy}^{0}(A)$ and high-field $%
\rho _{xy}^{0}>\rho _{xy}^{0}(E)$\ insulators. Note that in the case
illustrated by Fig.2, $i=1$\ at filling factor $\nu =2$ and $i=0$ at $\nu
=4,6...$.

Thus, the filling factors $\nu $ do not directly determine the positions of
the QH phases on the magnetic field axis, so $i\neq \nu $ at integer values
of $\nu $ if $\omega _{c}\tau \lesssim 1$. Sakr {\em et al.} \cite{Krav}
treat the shallow minima at $\nu =2,3,4,6$ with values of $\rho
_{xx}>0.3h/e^{2}$ as QH minima of the diagonal resistivity $\rho _{xx}$ with 
$i=\nu $. This interpretation of the experimental data seems questionable.

Note that equation(\ref{Ec}) does not describe exactly an electron system
with two different spin projections \cite{Khm2}, and the phase boundaries
are different from those plotted in Fig. 2. Even so, in the case of the spin
splitting energy smaller than $\hbar \omega _{c}$, the topology of the phase
diagram should remain the same.

In summary, the existence of the insulating phase between the quantum Hall
phases observed in Ref.\cite{Krav,Pud1,Pud6,Fang} is quite consistent with
the global phase diagram\cite{KLZ} for the quantum Hall effect. It is
possible owing to the Shubnikov-de Haas oscillations of ''bare'' diagonal
resistivity $\rho _{xx}^{0}$. The experimental results do not produce
evidence in favor of direct transitions from the insulating to quantum Hall
phases with large $i$.

I would like to thank S. I. Dorozhkin, V. M. Pudalov and V. N. Zverev for
helpful discussions. This work is supported by RFBR, PICS-RFBR and INTAS.

\end{document}